\documentclass[aps,prl,showpacs,preprintnumbers,twocolumn,superscriptaddress]{revtex4-2}
\usepackage{amsmath,amssymb}
\usepackage{mathtools}
\usepackage{physics}
\usepackage{bm}
\usepackage{tipa}
\usepackage{upgreek}
\usepackage{comment}
\usepackage{mathrsfs}
\usepackage{graphicx}
\usepackage{braket}
\usepackage{enumitem}
\usepackage{mathbbol}
\usepackage{booktabs}
\usepackage{gensymb}
\usepackage[normalem]{ulem}
\usepackage{color}
\usepackage[colorlinks,bookmarks=true,citecolor=blue,linkcolor=red,urlcolor=blue]{hyperref}
\usepackage{hyperref}

\usepackage{pifont}

\begin{document}

\title{Minimal Hubbard models of maximal Hilbert space fragmentation} 

\author{Yves H. Kwan}
	\affiliation{Princeton Center for Theoretical Science, Princeton University, Princeton NJ 08544, USA}
\author{Patrick H. Wilhelm}	
    \affiliation{Institut für Theoretische Physik, Universität Innsbruck, A-6020 Innsbruck, Austria}
\author{Sounak Biswas}
	\affiliation{Institut f\"ur Theoretische Physik und Astrophysik, Universit\"at W\"urzburg, 97074 W\"urzburg, Germany}
\author{S.A. Parameswaran}
	\affiliation{Rudolf Peierls Centre for Theoretical Physics, Parks Road, Oxford, OX1 3PU, UK}
	
\begin{abstract}
    We show that Hubbard models with
    nearest-neighbor hopping and a nearest-neighbor hardcore constraint exhibit `maximal' 
    Hilbert space fragmentation in many lattices of arbitrary dimension $d$. Focusing on the $d=1$ rhombus chain and the $d=2$ Lieb lattice, we demonstrate that the fragmentation is \emph{strong} for all fillings in the thermodynamic limit, and explicitly construct all emergent integrals of motion, which include an extensive set of higher-form symmetries. Blockades consisting of frozen particles partition the system in real space, leading to 
    anomalous dynamics. Our results are potentially relevant to optical lattices of
    dipolar and Rydberg-dressed atoms.
\end{abstract}

\maketitle

\textit{Introduction}.---Thermalization in closed quantum many-body systems has recently garnered increasing attention, partly spurred by advances in experimental platforms and quantum simulators~\cite{Kinoshita2006cradle,Gring2012prethermal,Trotzky2012relaxation,Schreiber2015mbl,Choi2016mbl,Kaufman2016thermalization,Smith2016mbl,Bernien2017_51,Kucsko2018dipolar,Morong2021stark,Scherg2021tilted}. While many-body localized~\cite{Nandkishore2015mblreview,Abanin2019mblreview} and integrable~\cite{Rigol2007integrable} systems have been long known as exceptions that break ergodicity, Hilbert space fragmentation~\cite{Moudgalya2022review} has emerged as a new route to violating  the eigenstate thermalization hypothesis (ETH)~\cite{Deutsch1991closed,Srednicki1994ETH,Rigol2008thermalization}. Here, additional symmetries or constraints cause the Hamiltonian or time evolution operator to fracture into exponentially many dynamically disconnected blocks. As a result, even within a symmetry sector, such systems exhibit anomalous thermalization. The earliest examples involved mobility restrictions due to the combination of charge and dipole conservation~\cite{Sala2020dipole,Khemani2020shattering,Moudgalya2021krylov,Pai2019fractonic}, but fragmentation has been subsequently uncovered in a wide variety of settings~\cite{Moudgalya2022commutant,DeTomasi2019Hubbard,Rakovszky2020tJ,Morningstar2020freezing,Chen2021emergent,Yang2020confinement,Bastianello2022tilted,Mukherjee2021constraint,Mukherjee2021minimal,Lee2021frustration,Hahn2021information,Li2021polar,Zadnik2021XXZ1,Zadnik2021XXZ2,Pozsgay2021XXZ,Khudorozhkov2022subsystem,Langlett2021fredkin,Richter2022motzkin,Brighi2022East,Yan2022height,Yoshinaga2022Ising,Hart2022Ising,Lian2023breakdown,Karpov2021gauge,Chakraborty2022spectral,Chakraborty2022transition,Frey2022Hubbard,stephen2022robust}.

Kinematic constraints often arise naturally from strong correlations. For example, the $t$-$J$ model that forbids doublon occupancy 
is the strong-coupling 
limit of the Hubbard model,
while the PXP model emerges within the blockaded subspace in Rydberg arrays. The low-energy properties of these and related Hamiltonians have been studied as settings for correlated and topological phenomena, and various one-dimensional versions have been shown to exhibit fragmentation and quantum many-body scars~\cite{DeTomasi2019Hubbard,Frey2022Hubbard,Li2021cluster,Rakovszky2020tJ,Moudgalya2021krylov,Mukherjee2021constraint,Mukherjee2021minimal,Chen2021emergent}.

\begin{figure}[t]
    \centering
    \includegraphics[width=0.9\columnwidth]{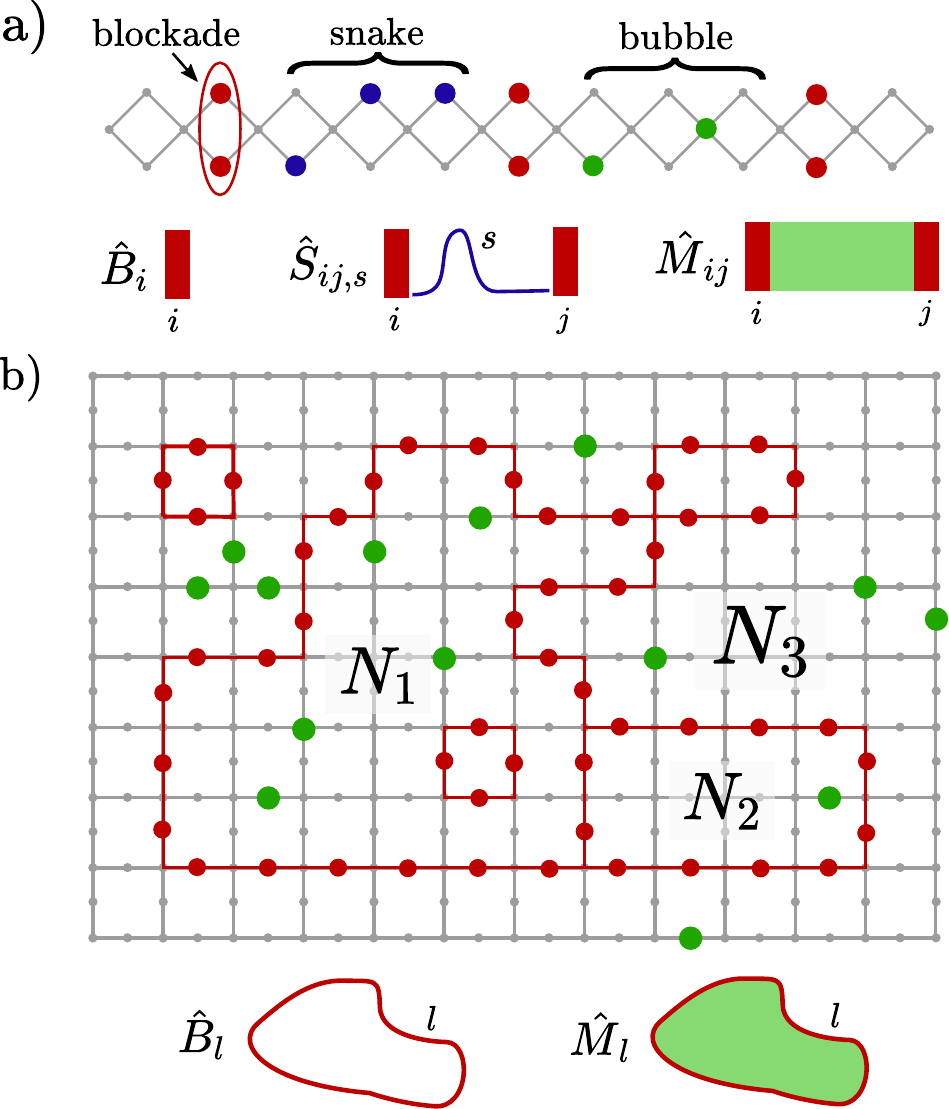}
    \caption{\textbf{Fragmention in extended hardcore Hubbard models.} a) Blockades in the rhombus chain are formed by two particles (red) on opposite corners, and demarcate frozen `snakes' (blue) and bubbles of mobile particles (green). b) Blockades on the Lieb lattice are due to closed frozen loops (red) of occupied particles on Lieb sites, which create bubbles with emergent subsystem charges $N_1,N_2,N_3$. For both lattices,  commutant algebra generators are schematically shown.}
    \label{fig:rhombus_Lieb}
\end{figure}

In this paper, we study Hilbert space fragmentation in spinless extended Hubbard models of strongly-interacting fermions or bosons with short-range hopping on general lattices. In the limit of infinite nearest-neighbor interactions $V_1$, their effects are encoded in kinematic constraints that forbid nearest-neighbor occupancy of particles. Focusing on two representative lattices, the rhombus chain and the Lieb lattice, we explicitly construct all emergent integrals of motion and demonstrate the existence of strong fragmentation for all fillings through analytical and numerical arguments. The non-trivial physics arises from an intuitive real-space picture of blockades which partition the system into disconnected subregions. In contrast to many previous studies of fragmentation, our model has a single global $U_c(1)$ symmetry corresponding to particle conservation, and is relevant for lattices in arbitrary dimensions $d\geq 1$. We discuss consequences for anomalous dynamics and thermalization,  generalizations to other lattices and settings, and assess the suitability of existing platforms involving dipolar or Rydberg-dressed atoms
as  settings for such phenomena.

\textit{Model}.---We consider a model of $N_p$ interacting spinless fermions or bosons on the $N_s$ sites of a graph in arbitrary dimensions. The filling factor is defined as $\nu=N_p/N_s$. Two key requirements are the restriction to nearest-neighbor (n.n.) hopping, and an extended hardcore constraint that prevents two particles from occupying the same or adjacent sites. The Hamiltonian is then
\begin{equation}\label{eq:H}
    \hat{H}=\sum_i w_i \hat{d}^\dagger_i \hat{d}^{\phantom\dagger}_i-\sum_{\langle{i,j}\rangle}t_{ij}\hat{d}^\dagger_i\hat{d}_j+\sum_{ij}V_{ij}\hat{n}_i\hat{n}_j,
\end{equation}
where $\hat{n}=\hat{c}^\dagger_i\hat{c}_i$, and $\hat{d}^\dagger_{i}=\hat{c}^\dagger_{i}\prod_{\langle j,i\rangle}(1-\hat{n}_j)$ is the dressed creation operator which acts within the hardcore subspace. The interaction $V_{ij}$ can be of arbitrary range, and the general site-dependence of the parameters means that the only conventional symmetry is the $U_c(1)$  corresponding to the global conserved charge $\hat{N}=N_p$.

For many lattices and fillings, Eq.~\ref{eq:H} is \emph{fragmented} in the Fock basis, meaning that  the Hamiltonian retains a non-trivial block-diagonal structure even when restricted to a symmetry sector labelled by $N_p$. In other words, the symmetry-resolved Hilbert subspace of dimension $D_\text{sym}$ decomposes into smaller Krylov fragments $\mathcal{K}_\alpha$ of size $\text{dim}(\mathcal{K}_\alpha)$, each of which are closed under Hamiltonian time evolution. The origin of this phenomenon lies in frozen subsets of particles which are completely immobile due to the hardcore constraint (e.g.~red pairs in Fig.~\ref{fig:rhombus_Lieb}a). In many cases, these frozen regions are \emph{blockades} that do not allow transmission of particles, and hence partition the system into disconnected \emph{bubbles} each of which has a conserved subsystem charge (see Fig.~\ref{fig:rhombus_Lieb}b). Each bubble can contain mobile particles and further nested blockades. Fragments with distinct blockade networks and bubble charges cannot be connected by $\hat{H}$. As shown later, the fragmentation structure is characterized by a hierarchy of emergent higher-form symmetries.

\textit{Rhombus chain}.---Our first example lattice is the $d=1$ rhombus chain (RC) with $L$ unit cells, each of which contains two corner sites `$\land,\lor$' and one junction site `$\times$' (Fig.~\ref{fig:rhombus_Lieb}a). The maximum allowed filling is $\nu=2/3$. Two particles on opposite corner sites comprise a blockade. Between two consecutive blockades, there is either a bubble region of mobile particles, or a string of frozen particles dubbed a snake. Within a global charge sector, a Krylov fragment can be uniquely identified by specifying the blockade positions, and the intervening snake configurations and bubble charges.

This physics is succinctly captured by the commutant algebra $\mathcal{C}$, which Ref.~\cite{Moudgalya2022commutant} used to characterize the
fragmented structure shared by families of models. $\mathcal{C}$ consists of all operators which commute with every term of $\hat{H}$. Besides the trivial identity operator $\mathbb{1}$, the global $U_c(1)$ provides the usual symmetry algebra consisting of functionally independent powers of $\hat{N}$: there are $\mathcal{O}(L)$ such combinations, consistent with a continuous global symmetry. We now turn to the novel integrals of motion (IOMs) (Fig.~\ref{fig:rhombus_Lieb}a, bottom). For simplicity, we ignore boundary effects below. Individual blockades can be diagnosed by a set of projectors $\hat{B}_i=\hat{n}_{i\wedge}\hat{n}_{i\lor}$  onto pairs of occupied rhombus corners. Snakes are detected by operators $\hat{S}_{ij,s}=\hat{B}_i\hat{B}_j\prod_{i<k<j}\hat{n}_{k,s_k}(1-\hat{n}_{k,\bar{s}_k})$, where $s$ is a string of $\land,\lor$ of length $j-i-1$ and $\bar{s}_k$ indicates the opposite corner. Similarly, charges within bubbles are denoted $\hat{M}_{ij}=\hat{B}_i\hat{B}_j\sum_{i<k<j}\hat{n}_k$. Note that $\hat{S}_{ij,s},\hat{M}_{ij}$ are non-locally conditioned operators which are only active if the blockades are present. A generating set for the commutant algebra for the RC is
\begin{equation}\label{eq:genC_rhombus}
    \text{gen}(\mathcal{C})=\{\mathbb{1},\hat{N},\hat{P}_\times,\{\hat{B}_i\},\{\hat{S}_{ij,s}\},\{\hat{M}_{ij}\}\}
\end{equation}
where we have also included the projector $\hat{P}_\times$ onto the frozen $\nu=1/3$ state with all junction sites filled. Since $\mathcal{C}$ is Abelian, the fragmentation is `classical' and hence manifest in a product basis (the Fock states)~\cite{Moudgalya2022commutant}. 
$\text{dim}(\mathcal{C})$ gives the total number of Krylov subspaces across all $N_p$ sectors. 
The algebra generated by Eq.~\ref{eq:genC_rhombus} provides the entire structure of emergent IOMs. Evidently, this grows exponentially with system size as there are at least $2^L$ individual snake operators alone. 

\begin{figure*}[t!]
    \centering
    \includegraphics[width=1\linewidth]{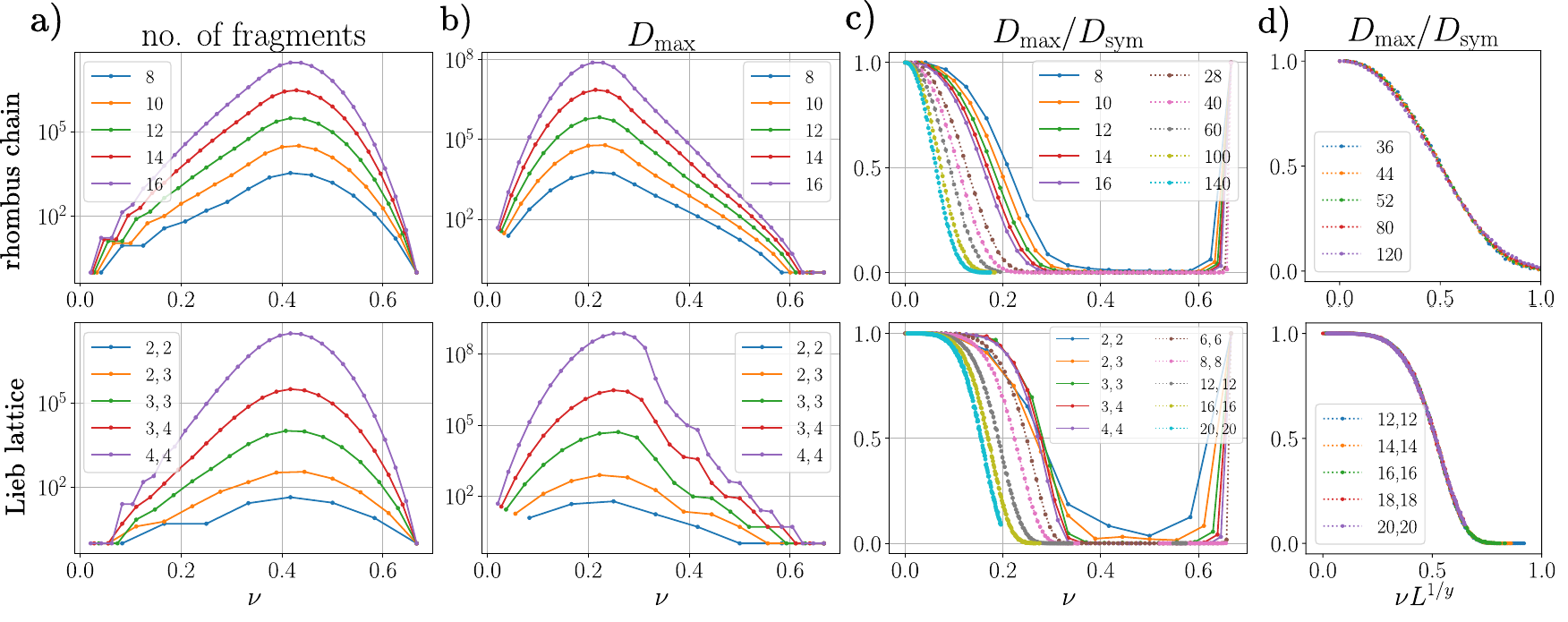}
    \caption{\textbf{Exact enumeration (EE) and Monte Carlo (MC) sampling of fragments.} Top (bottom) row indicates results on $L$ ($N_x\times N_y$) unit cells of the rhombus chain (Lieb lattice) with periodic boundary  conditions. a) Number of fragments in EE for each filling factor $\nu$. b) Dimension of largest Krylov fragment $D_\text{max}$ in EE. c) Ratio of $D_\text{max}$ to the dimension of the total symmetry sector $D_\text{sym}$. There is a finite-size drift $D_\text{max}/D_\text{sym}\rightarrow0$, suggesting that the system is strongly-fragmented for all fillings in the thermodynamic limit. Solid (dotted) lines refer to EE (MC) data. All MC data points include at least 3000 samples. d) Scaling collapse of MC data for $D_\text{max}/D_\text{sym}$ as a function of $(\nu-\nu_c) L^{1/y}$ with fixed $\nu_c=0$, and fitted exponents $y=2.4(1)$ ($y=2.5(1)$) for the rhombus chain (Lieb lattice). 
    \label{fig:EEMC}}
\end{figure*}

\textit{Lieb lattice}.---We now turn to the square Lieb lattice (SLL) of $N$ unit cells, which provides the simplest example of extensive fragmentation~\footnote{With the benefit of hindsight, the solvable limit of the Emery model on a similar lattice in Ref.~\onlinecite{kivelsonemerymodel} could be viewed as an early example of {\it subextensive} fragmentation.} in $d=2$. Each unit cell contains one square lattice site and two `Lieb' sites on the links, and the maximum filling consistent with the hardcore constraint is $\nu=2/3$. Blockades can be introduced by drawing arbitrary closed loops on the links of the square lattice, and placing particles on all of the Lieb sites that are traversed (Fig.~\ref{fig:rhombus_Lieb}b). Primitive blockades, which cannot be decomposed into smaller ones, are equivalent to self-avoiding polygons on the square lattice (a similar  blockade-loop-structure also occurs in height-conserving dimer models~\cite{Yan2022height}). The smallest  blockade is a single plaquette with four Lieb particles. Note that unlike in $O(n)$ loop models, loops are allowed to share edges and intersect. The configuration of loops divides the system into bubbles with conserved subsystem charges.

As in the RC, we can construct extended operators which 
diagnose the Krylov fragments. Neglecting boundary effects for simplicity, we define loop blockade projectors $\hat{B}_l=\prod_{\text{Lieb } i\in l}\hat{n}_i$, where the product is over Lieb sites on  loop $l$. For each loop, we also define the membrane operator $\hat{M}_l=\hat{B}_l\sum_{i\in b(l)}\hat{n}_i$ which counts the charge inside a bubble $b(l)$ conditioned on the existence of a surrounding blockade.
These operators have support on a line contour and are  examples of one-form symmetries (Fig.~\ref{fig:rhombus_Lieb}b), though they are only non-vanishing on Krylov subspaces that contain the corresponding blockade.
The commutant algebra for the SLL is generated by
\begin{equation}\label{eq:genC_Lieb}
    \text{gen}(\mathcal{C})=\{\mathbb{1},\hat{N},\hat{P}_\text{sq.},\{\hat{B}_l\},\{\hat{M}_{l}\}\}
\end{equation}
where $\hat{P}_\text{sq.}$ is the projector onto the frozen $\nu=1/3$ state with all square lattice sites filled. The number of fragments is exponential (since, e.g.~counting products of single-plaquette loop operators alone gives $2^N$ possibilities). With periodic boundary conditions (PBCs), there are also non-contractible loop operators. The associated membrane operators count the charge between two such loops with identical winding numbers.

\textit{Strong fragmentation}.---While we have demonstrated an exponential number of fragments for the RC and SLL, their dimensions and distribution across different $N_p$ sectors is not yet clear.
To understand the extent of ergodicity breaking, it is useful to contrast
{\it strong} and {\it weak} fragmentation~\cite{Sala2020dipole,Khemani2020shattering}. Within a conventional symmetry sector (i.e.~filling factor $\nu$ in the thermodynamic limit), a system is strongly fragmented if the size of the largest fragment $D_\text{max}=\text{max}_\alpha\{\dim(\mathcal{K}_\alpha)\}$ is a vanishing fraction of $D_\text{sym}$ as the number of unit cells $N\rightarrow \infty$. On the other hand, $D_\text{max}/D_\text{sym}\rightarrow1$ for weak fragmentation, where typical states are expected to look thermal~\cite{Moudgalya2022review}. We first tackle the simpler question of frozen Fock states. It turns out that the problem of enumerating either frozen or general hardcore configurations can be mapped onto a \emph{local} vertex model~\cite{SupMat}, which can be efficiently simulated by transfer matrices for the RC and tensor renormalization group for the SLL~\cite{Levin2007TRG,Gu2009TEFR}. We find that while the number of frozen states grows exponentially, the frozen fraction still vanishes
as $N\to\infty$ for almost all fillings~\cite{SupMat}.

Distinguishing between weak and strong fragmentation is more challenging since it involves a complex interplay between the extensive entropy of mobile regions and the configurational entropy of blockade networks. It is natural that strong fragmentation, if it exists, occurs at larger fillings where there are likely more blockades and smaller bubbles. 
In fact, we demonstrate that there is strong fragmentation as $\nu\rightarrow0$. We first note that if the system were weakly fragmented, then $D_\text{max}$
must correspond to the fragment with no blockades ($D_\text{no bl.}$), since otherwise, spatially translating all blockade positions generates  distinct equal-size fragments, whose number necessarily grows with increasing $N$ and dilute fillings. Since $D_{\rm sym}$ and $D_{\text{no bl.}}$ are given by partition functions of classical hardcore models,
we can perform low-filling cluster expansions~\cite{Mayer,FriedliVelenik}
    in $\nu$, and find  
    $D_{\text{no bl.}}/D_{\rm sym} \sim \exp(-c N \nu ^s)$ as $\nu\to0$,
    with a lattice-dependent constant $c$ and an exponent $s$ set by  the number of particles in the the smallest blockade~\cite{SupMat}.
 This implies strong fragmentation at all low fillings in the thermodynamic limit. The exponential is reminiscent of the finite-size scaling form for  an incipient phase transition at $\nu = 0^+$,   with a correlation volume growing as $\sim \nu^{-s}$.

To understand more quantitatively the distribution of fragments and the strong/weak phase diagram, we perform an exact enumeration (EE) of fragments. This involves a depth-first search algorithm which analyzes all the states in a symmetry sector and groups them based on if they are connected by $\hat{H}$~\cite{SupMat,Zhang2003}.
As shown in Fig.~\ref{fig:EEMC} for PBCs, the number of fragments is exponential in system volume for most fillings. $D_\text{max}/D_\text{sym}$ for the RC shows clear size-dependence, suggesting a drift of the weak-to-strong crossover $\nu_c$ towards lower fillings, but the data for SLL are more difficult to interpret due to the small systems and finite-size effects.

While EE provides full information on the fragments for finite systems, it is quickly limited by the exponential growth of the Hilbert space. Therefore, we use Monte Carlo (MC) sampling~\cite{SupMat} to follow the $\nu$-dependence of the weak-strong crossover to larger systems. We use the vertex models discussed earlier to generate a large set of unbiased samples. Each sample is analyzed and assigned to the appropriate fragment. From this, we can compute $D_\text{max}/D_\text{sym}$
as the proportion of samples in the largest accumulated fragment. As shown in Fig.~\ref{fig:EEMC}c, the data show a clear drift of the weak-strong crossover to lower fillings. A scaling collapse strongly suggests that the critical filling $\nu_c=0$ in the thermodynamic limit~(Fig.~\ref{fig:EEMC}d), implying strong fragmentation for all fillings, though the best fit exponents indicate the presence of significant finite-size effects.

\textit{Thermalization and localization}.---Strong fragmentation leads to violation of the weak form of ETH within each global particle number sector~\cite{Moudgalya2022review}. Local observables such as the density can differ dramatically for nearby states in the middle of the spectrum. For instance, the site $i$ may be part of a blockade in one eigenstate so that $\langle \hat{n}_i\rangle=1$, but part of a dilute thermal bubble for another eigenstate at the same energy density. The bipartite entanglement entropy vanishes in fragments where the entanglement cut traverses a blockade whose width is sufficiently larger than the interaction range, and is generally much lower than the Page value expected for chaotic systems. The level spacings 
are expected to follow a Poisson distribution because of the absence of level repulsion between different fragments.

\begin{figure}[t]
    \centering
    \includegraphics[width=0.9\columnwidth]{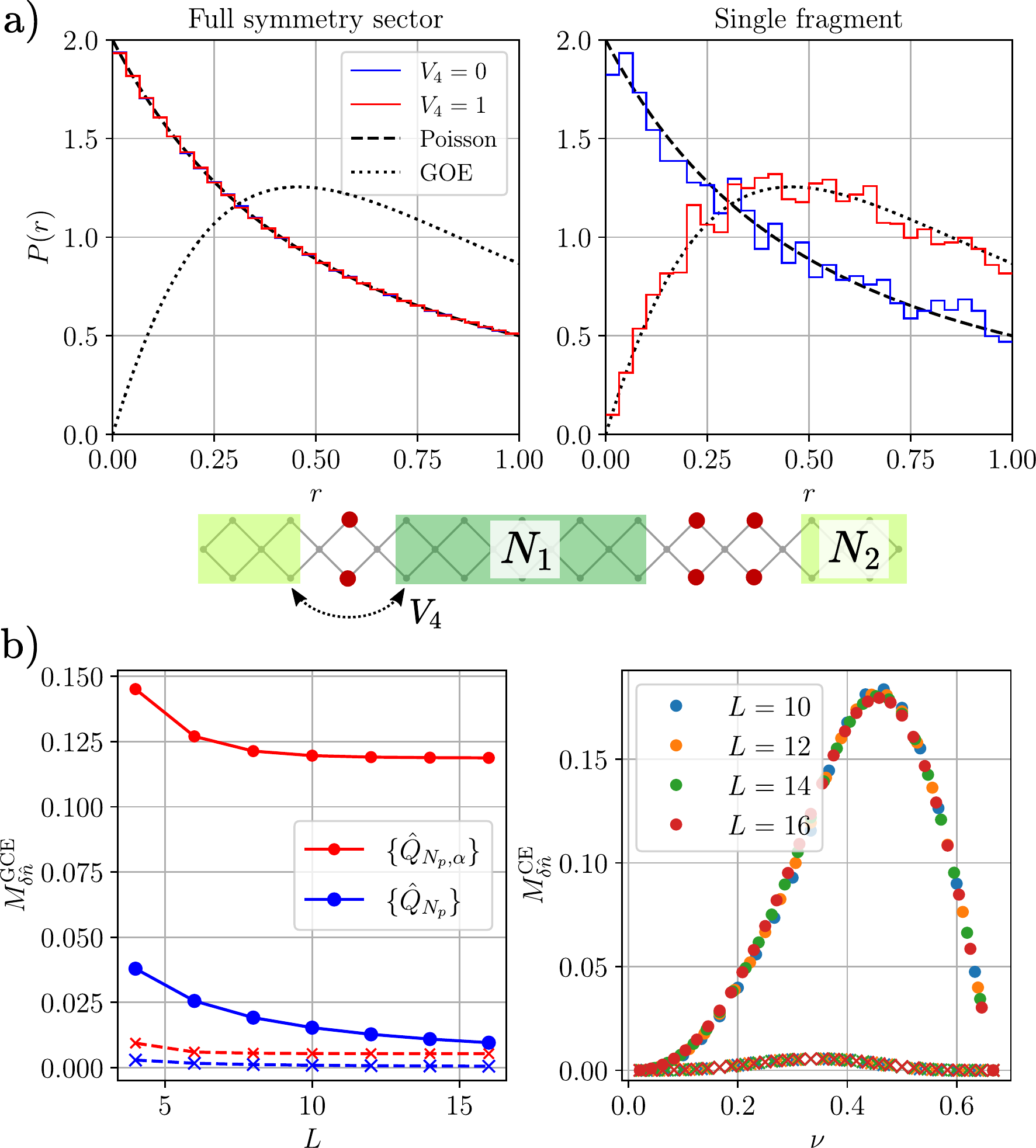}
    \caption{\textbf{Thermalization and dynamics in the rhombus chain.} a) Left: Distribution $P(r)$ of  level spacing ratios $r$ for  symmetry sector $L=12,N_p=13$, showing Poisson statistics. Right: $P(r)$ for the two-bubble fragment shown at bottom with emergent charges $N_1=4,N_2=3$, showing the change from Poisson to GOE statistics on including horizontal-range-4 ($V_4$) interactions. b) Mazur bounds for the dynamical autocorrelator of $\delta \hat{n}_i=\hat{n}_i-\langle n\rangle_i$. Left: Red (blue) indicates bounds in the grand canonical ensemble, taking into account fragmentation (only $N_p$ conservation). Circles (crosses) indicate the $\land$ ($\times$) sublattice. Right: Bounds obtained using fragmentation in the canonical ensemble.}
    \label{fig:thermalization}
\end{figure}

Despite the ergodicity-breaking outlined above, one may ask whether individual Krylov subspaces are themselves thermal~\cite{Moudgalya2021krylov}. The answer depends on the fragment and the range of interactions. Consider the fragment on the Lieb lattice that is represented in Fig.~\ref{fig:rhombus_Lieb}b which has bubbles $b_l$ with corresponding conserved charges $N_l$. For $V_{ij}=0$ (i.e.~the only interactions are the hardcore constraints), the bubbles do not interact with each other. We therefore have isolated subsystems that can realize various physics depending on their disorder, dimensionality, and charge. Sufficiently disordered 1d bubbles may be Anderson or many-body localized, while large 2d bubbles are expected to locally satisfy ETH, but the fragment overall fails to equilibrate. As interactions $V_{ij}$ of progressively longer range are introduced, the bubbles begin to interact and thermalize each other, and only bubbles with sufficiently thick `shielding' boundaries remain isolated. For infinite-range interactions, we expect Krylov-restricted thermalization to occur for all fragments. We demonstrate this in the RC by using exact diagonalization (ED) to compute the distribution of level spacings $r=\frac{\text{min}(s_n,s_{n-1})}{\text{max}(s_n,s_{n-1})}$ \cite{Oganesyan2007,Atas2013}, where $s_n=E_{n+1}-E_n$. Fig.~\ref{fig:thermalization}a shows that a typical fixed-$N_p$ sector exhibits Poisson statistics. However, an individual fragment in that sector can cross over from Poisson to Gaussian orthgonal ensemble (GOE) statistics as $V_4$ is increased.

The presence of blockades influences the nature of transport and dynamics in these models. While long-range interactions permit bubbles to transmit information, particle exchange remains forbidden. The propagator $\langle \hat{c}^\dagger_i \hat{c}^{\phantom{\dagger}}_j\rangle$ vanishes in fragments where $i,j$ belong to different bubbles.
Even if there is a percolating mobile region, other bubbles act as scatterers, possibly with internal degrees of freedom, that may hinder charge transport. The fragmentation also contributes to a lower (Mazur) bound $M_{\delta\hat{n}}$ on the infinite-temperature dynamical charge autocorrelation function~\cite{Moudgalya2022commutant,SupMat}
\begin{equation}\lim_{\tau\rightarrow\infty}\frac{1}{\tau}\int_0^\tau dt \langle \delta\hat{n}_i(t) \delta\hat{n}_i(0)\rangle\geq M_{\delta\hat{n}},
\end{equation}
where $\delta\hat{n}_i=\hat{n}_i-\langle \hat{n}_i\rangle$ is traceless. This can be extracted using EE, as detailed in Ref.~\cite{SupMat}. As shown for the RC in Fig.~\ref{fig:thermalization}b (and for the SLL in~\cite{SupMat}), in the grand canonical ensemble, the global $U_c(1)$ symmetry only constrains $M_{\delta\hat{n}}\sim1/L$, but the bound saturates to a finite value when accounting for the fragmented structure. Similar behavior occurs within a fixed-$N_p$ sector, where the only contributions to $M_{\delta\hat{n}}$ arise from the emergent symmetries. The bound is
stronger for sublattices that participate in blockades, since such charges cannot relax at all.

\textit{Extensions}.---We can consider relaxing the extended hardcore constraint while taking $V_1\rightarrow\infty$.  In this case, the number of occupied n.n.~bonds is a conserved quantity, which underlies the distinct form of fragmentation studied previously in 1d hardcore Hubbard chains~\cite{DeTomasi2019Hubbard,Frey2022Hubbard}. A pair of n.n.~particles remains frozen unless a third particle approaches and enables a resonant hop (if $V_1$ is uniform), but the blockades shown in Fig.~\ref{fig:rhombus_Lieb} remain frozen and impenetrable. Similarly, we can relax the onsite constraint and reintroduce spins, while preserving the blockade physics. For finite n.n.~interactions, strictly speaking the fragmentation disappears completely as now it is possible for particles to pass through a blockade, or for the blockade to melt. However for large $V_1$, the fragmentation is still expected to leave an imprint on dynamics since the effective n.n.n.~hopping that is generated is suppressed as $\sim t^2/V_1$. Thicker blockades will take longer to break down. Similar comments apply to longer-range hoppings directly present in the Hamiltonian: while typically exponentially suppressed, they make the emergent conservation laws  only approximate. In these cases, the dynamics may show features of fragmentation  at short times, but these are eventually washed out as $t\to\infty$.

In 2d, strong fragmentation is expected to be possible in other graphs which allow for finite blockades, such as the dice lattice and P3 Penrose tilings~\cite{SupMat}. In fact, any 2d graph can be converted into one that hosts blockade loops by inserting `Lieb' sites in the middle of every edge. Other lattices, such as the square and honeycomb lattices, have blockades which must span the system size~\cite{SupMat}, and hence are expected to be weakly fragmented for all $\nu$ (though they may exhibit a finite-size strong-weak crossover with its own scaling properties~\cite{Hart2022Ising}). Higher-dimensional generalizations are possible, where the formation of bubbles with subregion charge conservation requires closed $(d-1)$-dimensional blockade hypersurfaces.

We propose that the physics here can be probed in existing experimental platforms that realize extended variants of the (Bose)-Hubbard model~\cite{Lewenstein2007ultracold,lewenstein2012book,Dutta2015nonstandard}. The non-trivial requirement is the ability to generate the desired lattices and engineer sufficiently strong onsite and n.n.~interactions to effectively enforce the extended hardcore constraint. Ideally, the interaction falls off rapidly beyond n.n.~range so that dynamical process within the bubbles are not energetically frozen out. Interference of laser beams in optical lattices of ultracold gases has been leveraged in experiments to realize various potentials such as the Lieb lattice~\cite{Taie2015Lieb,Ozawa2017lieb,Shen2010lieb,Apaja2010lieb,Flannigan2021lieb}. Dipolar atoms or molecules with power-law decaying $V_{ij}$ can provide the required hierarchy of couplings when combined with tuning of the hopping strengths~\cite{Trefzger2011dipolar,Lahaye2009dipolar,CapogrossoSansone2010polar,Baranov2012dipolar,Bohn2017molecules,Moses2016polar,Baier2016extended,Chomaz2022dipolar}. Disorder-free localization and fragmentation has previously been proposed in such a setting for the 1d chain~\cite{Li2021polar}. Similarly, Rydberg-dressed atoms inherit a strong repulsion within a tunable blockade radius and a faster long-range $r^{-6}$ falloff~\cite{Henkel2010rydberg,Pupillo2010rydberg,Johnson2010rydberg,Schau2015crystallization,Browaeys2020rydberg,GuardadoSanchez2021quench}. (Such systems have been proposed to realize a topological Mott insulator on the Lieb lattice~\cite{Dauphin2016mott}.) The temperature can be comparable to the couplings, as long as it is much less than the n.n.~interaction $V_1$.

\textit{Conclusions and outlook}.---We have shown that the combination of nearest-neighbour hoppings and the simplest possible extended hardcore constraint in Hubbard models induces Hilbert space fragmentation in many lattices of different dimensions. The presence of frozen blockade regions inhibits charge transfer between emergent bubbles, and leads to an exponential hierarchy of integrals of motion. The necessary ingredients for realizing such ergodicity-breaking are already present in current experimental platforms such as ultracold atomic gases in optical lattices. While we have focused on two examples that exhibit strong fragmentation for all fillings, it would be interesting to study the extent of fragmentation and the scaling behavior of $D_\text{max}/D_\text{sym}$ in other physically relevant lattices.
For rational couplings, investigating \emph{quantum} fragmentation~\cite{Moudgalya2022commutant} via integer polynomial factorization~\cite{Regnault2022integer} is another interesting direction. We leave to future work the question of how the anomalous dynamics in these models survives the inclusion of perturbations that lift the fragmentation structure.

\begin{acknowledgements}
\textit{Acknowledgements}.---We thank Andrei Bernevig, Sarang Gopalakrishnan, Vedika Khemani, Steve Kivelson, Abhinav Prem, and Nicolas Regnault for useful discussions. We acknowledge support from  the European Research Council (ERC) under the European Union Horizon 2020 Research and Innovation Programme (Grant Agreement Nos.~804213-TMCS),  EPSRC grant EP/S020527/1, and the Austrian Science Fund FWF within the DK-ALM (W1259-N27). Statement of compliance with EPSRC policy framework
on research data: This publication is theoretical work
that does not require supporting research data.    
\end{acknowledgements}
%


\newpage
\clearpage

\begin{appendix}
\onecolumngrid
	\begin{center}
		\textbf{\large --- Supplementary Material ---\\ Minimal Hubbard models of maximal Hilbert space fragmentation}\\
		\medskip
		\text{Yves H. Kwan, Patrick H. Wilhelm, Sounak Biswas, and S.A. Parameswaran}
	\end{center}

\section{Vertex model for the rhombus chain}

\begin{figure}[t]
    \centering
    \includegraphics[width=0.60\columnwidth]{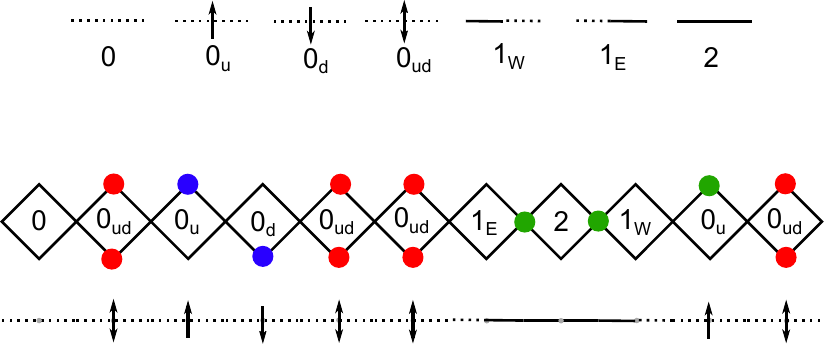}
    \caption{Top: Vertices in the 7-vertex model relevant for hardcore configurations on the rhombus chain. Only the first 4 vertices are kept for frozen configurations. Bottom: Example state in the particle language and its vertex representation. Red particles are blockades, blue particles are part of frozen snakes, and green particles are mobile.}
    \label{fig:app_rhombus_vertex}
\end{figure}

Hardcore configurations on the rhombus chain (RC) with $L$ unit cells can be mapped to vertex configurations of a 7-vertex model on a chain of $L$ sites. As shown in Fig.~\ref{fig:app_rhombus_vertex}, each rhombus in the hardcore model is assigned to a site in the vertex model, and the pattern of particle occupancy determines the vertex. A vertex has two legs, which can have links (a solid leg). Legs without links are referred to as empty legs.  Note that all $2^2$ combinations of links on the two legs are allowed. In addition to the linkless vertex (labelled `$0$'), there are three additional vertices ($0_\text{u},0_\text{d},0_\text{ud}$) with `decorations' but no links. Such vertices with all empty legs are referred to here as ghost vertices. The vertices obey adjacency rules depending on the links. For instance, $1_\text{E}$ can lie to immediately to the left of $2$, but not vice versa. To capture only frozen configurations (blockades and snakes), we restrict to a 4-vertex model with vertices $0,0_\text{u},0_\text{d},0_\text{ud}$. In addition, we impose additional adjacency rules such that $0_\text{u}$ and $0_\text{d}$ are not allowed to be adjacent to $0$. We note that while snakes and bubbles can be extended non-local objects, the relevant vertex model is \emph{local}.

We can make a simple Pauling-type estimate for the total number of hardcore configurations. For one sublattice (say A) of the vertex model, we can freely and independently choose the vertices, leading to $7^{L/2}$ possibilities. Since the 7 possible vertices in total have 4 links out of a possible maximum of 14, the legs on sublattice B have a $2/7$ chance of being solid, and $5/7$ chance of being empty. If a B vertex has two links (probability $4/49$) or one link (probability $20/49$), then its vertex is uniquely specified. If both legs are empty (probability $25/49$), there are four possible (ghost) vertices. This leads to a total hardcore configuration count of $(7\times(\frac{4}{49}+\frac{20}{49}+3\times\frac{25}{49}))^{L/2}\simeq 4.209^L$. Similarly, the number of hardcore configurations scales as $\simeq 3.082^L$. Both estimates are close to the numerical results.

\begin{figure}[t]
    \centering
    \includegraphics[width=0.60\columnwidth]{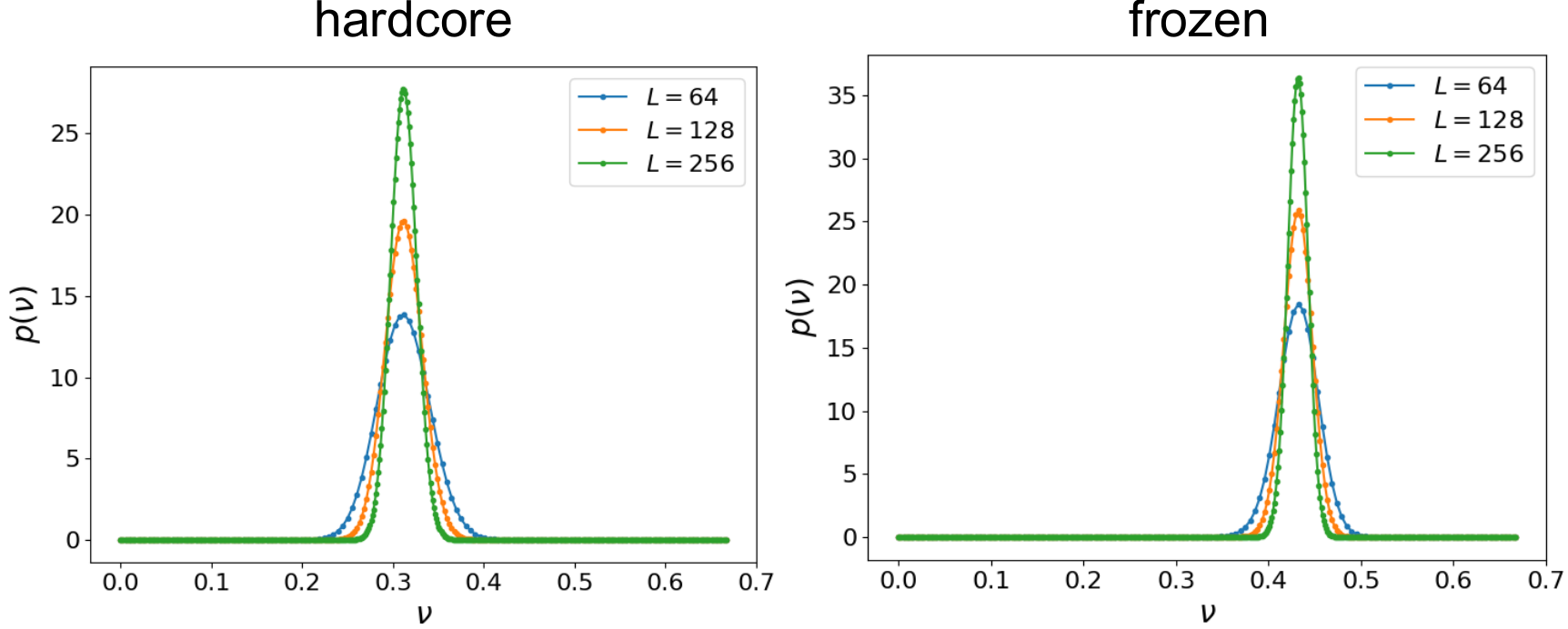}
    \caption{Probability distribution functions for hardcore and frozen configurations on the rhombus chain as a function of the filling $\nu$.}
    \label{fig:app_rhombus_transfer}
\end{figure}

Transfer matrices provide a simple method to extract the total number and distribution of frozen/hardcore configurations. The goal is to compute the complex moment generating function
\begin{equation}\label{eqn:app:Mk}
    M(k)=\sum_{c}e^{i\frac{2\pi kn(c)}{n_\text{max}+1}}=\text{Tr}\left(T(k)^L\right)
\end{equation}
where $n_\text{max}=2L$ is the maximum number of particles, $n(c)$ is the particle number corresponding to the configuration $c$, and $k=0,1,\ldots,n_\text{max}$. $c$ runs over the relevant set of configurations. $M(0)$ gives the total number of such configurations, and the inverse Fourier series provides the number distribution function as a function of the particle number $n$. For general hardcore configurations, the matrix $T(k)$ is given by
\begin{gather}
    T(k)=\exp\left[i\frac{2\pi k}{n_\text{max}+1}W(k)\right]A(k)\\ W(k)=\text{diag}[0,1,1,2,1/2,1/2,1],\quad A(k)=\begin{pmatrix}
    1&1&1&1&0&1&0\\
    1&1&1&1&0&1&0\\
    1&1&1&1&0&1&0\\
    1&1&1&1&0&1&0\\
    1&1&1&1&0&1&0\\
    0&0&0&0&1&0&1\\
    0&0&0&0&1&0&1
    \end{pmatrix}.
\end{gather}
$W(k)$ encodes the particle number carried by each vertex (taking into account double-counting), and $A(k)$ is the adjacency matrix that encodes which vertices are allowed to  be adjacent. The basis ordering is the same as in Fig.~\ref{fig:app_rhombus_vertex}. The corresponding matrices for the 4-vertex model for frozen configurations can be straightforwardly deduced. The probability distribution functions for various chain lengths are shown in Fig.~\ref{fig:app_rhombus_transfer}. As expected, the distributions narrow as $L$ increases.

The number of frozen states scales as $\sim3.17^L$ and peaks at $\nu\simeq 0.432$, while the number of hardcore states scales as $\sim 4.303^L$ and peaks at $\nu\simeq 0.311$

\section{Vertex model for the Lieb lattice}
\begin{figure}[t]
    \centering
    \includegraphics[width=0.46\columnwidth]{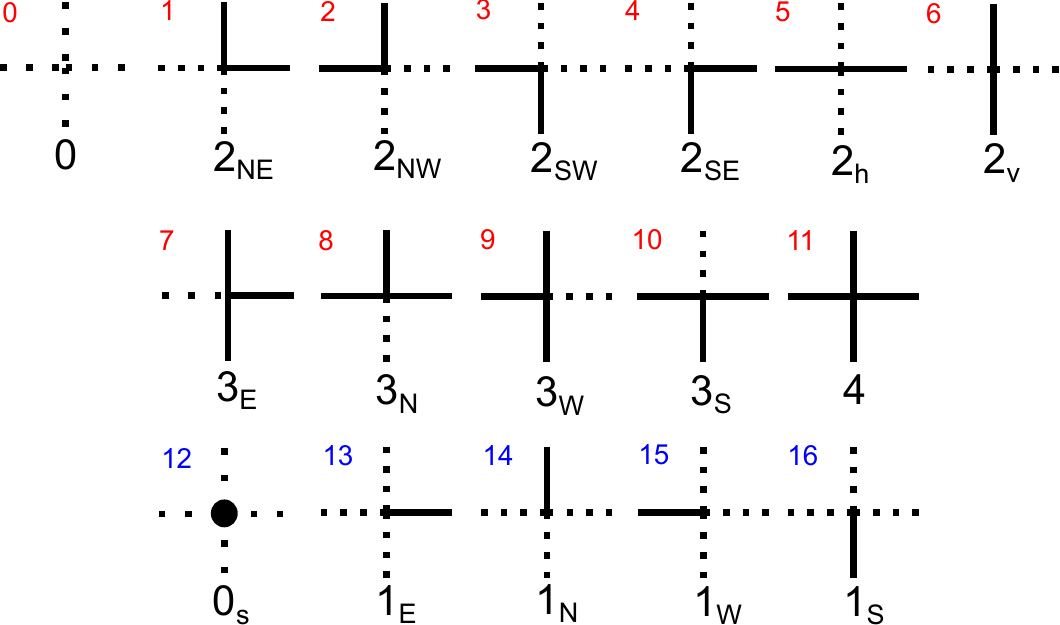}
    \caption{Vertices in the 17-vertex model relevant for hardcore configurations on the Lieb lattice. Only for the first 12 vertices are kept for frozen configurations. The vertex model has sites on each square lattice site of the Lieb lattice. The presence of a link on a leg indicates the presence of a particle on the corresponding Lieb site. The $0_\text{s}$ vertex indicates a particle on the corresponding square lattice site.}
    \label{fig:app_Lieb_vertex}
\end{figure}

Hardcore configurations on the Lieb lattice with $N_\text{UC}$ unit cells can be mapped to vertex configurations of a 17-vertex model on a square lattice with $N_\text{UC}$ sites. The vertices are shown in Fig.~\ref{fig:app_Lieb_vertex}. A link corresponds to an occupied Lieb site in the hardcore model, and the $0_\text{s}$ vertex captures square lattice particles. Note that all $2^4$ combinations of lilnks on the four legs are allowed. In addition to the linkless vertex $0$, there is one additional vertex $0_\text{s}$ with a site decoration---hence there are two ghost vertices. The adjacency rules are based on whether the legs match up. To capture only frozen configurations (i.e.~loop configurations), we restrict to a 12-vertex model with just the first 12 vertices. This excludes occupied square lattice sites, as well as `dangling bonds' in the link language. 

We can make a Pauling estimate of the entropy of hardcore configurations. On one sublattice of the vertex model, we can pick among $17^{N_{\text{UC}}/2}$ configurations independently. The other sublattice is guaranteed to be in a legal state. But if all links are empty, there is an additional choice of filling the square lattice site (i.e.~vertices $0$ and $0_\text{s}$). Since there is a $9/17$ chance for a link to be empty, the total count is $\left(17\cdot(1+\frac{9^4}{17^4})\right)^{N_{\text{UC}}/2}\simeq 4.282^{N_\text{UC}}$. For frozen configurations, we make a similar argument. On one sublattice, we the vertices independently leading to $12^{N_{\text{UC}}/2}$ configurations. There are in total 28 links (out of a possible maximum of 48) in the 12 possible vertices, so on the other sublattice, each leg has a $5/12$ chance of being empty and $7/12$ chance of have a link. Hence on each site of the other sublattice, there is a $1-4\cdot\frac{5^3\cdot7}{12^4}\simeq 0.831$ chance of it being in a legal state (the illegal ones are where there is only one link). This leads to a total count of $\left(12\cdot(1-4\cdot\frac{5^3\cdot7}{12^4})\right)^{N_{\text{UC}}/2}\simeq 3.158^{N_\text{UC}}$, where we have neglected the correction arising if all legs on the second sublattice are empty (then we can pick from two ghost vertices). Both estimates are close to the numerical results.

\begin{figure}
    \centering
    \includegraphics[width=.5\textwidth]{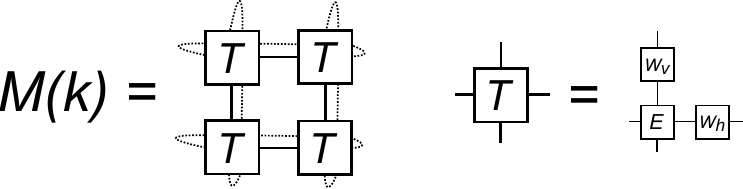}
    \caption{Schematic of tensor network for the Lieb lattice, shown for a $2\times2$ system with periodic boundary conditions.}
    \label{fig:app:tensors}
\end{figure}

Since we have classical local vertex models in 2d, we can use the tensor renormalization group (TRG) to efficiently compute the partition function~\cite{Levin2007TRG,Gu2009TEFR}. The approximation in TRG is that there is a `bond dimension' $\chi$ which controls the truncation error of the tensors.  Here, we discuss how to obtain the number and distribution (as function of particle number) of frozen states and hardcore states, using the 12- and 17-vertex models. The goal is again to compute the complex moment generating function in Eq.~\ref{eqn:app:Mk}, except now the trace is over a 2d tensor network as shown in Fig.~\ref{fig:app:tensors}. The tensor $E_{ijkl}=\epsilon_i\delta_{ijkl}$ captures the local contribution of each vertex to the phase in Eq.~\ref{eqn:app:Mk}, accounting for double-counting of links if necessary. The tensors $W_h,W_v$ encode the adjacency rules of the vertex model. The TRG consists of $S$ steps, which corresponds to a periodic Lieb lattice system of $2^S$ unit cells.  The maximum particle number in both 12- and 17-vertex models is $n_{\text{max}}=2^{S+1}$.

\begin{figure}
    \centering
    \includegraphics[width=.5\textwidth]{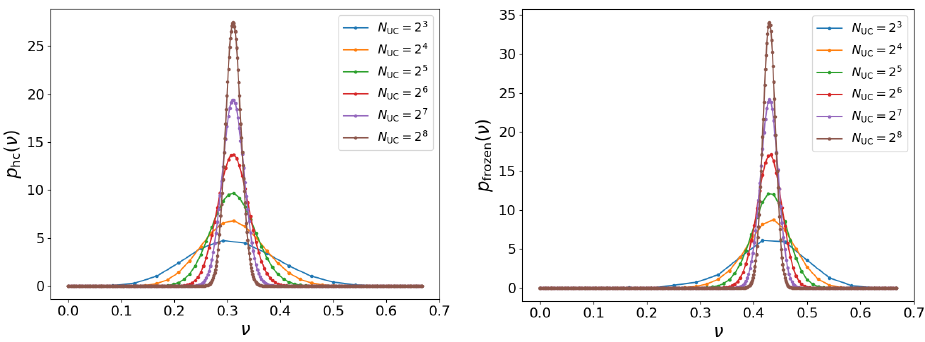}
    \caption{TRG results for the 17-vertex (hardcore) and 12-vertex (frozen) models for the Lieb lattice. $\chi=20$.}
    \label{fig:app:TRGresults}
\end{figure}

In Fig.~\ref{fig:app:TRGresults}, we show the TRG results for the distributions of frozen and hardcore configurations. Not surprisingly, the probability distributions get increasingly narrow as the system size increases. We find that frozen (hardcore) states scale as $\sim3.151^L$ ($\sim4.286^L$) and peak at $\nu=0.429$ ($\nu=0.311$)

\section{Monte Carlo sampling of hardcore configurations}

In this section, we discuss the implementation of Monte Carlo (MC) sampling for vertex models corresponding to systems such as the Lieb lattice and rhombus chain. By sampling the set of hardcore configurations in the right way, one can estimate, for a given filling, the ratio $D_{\text{max}}/D_{\text{sym}}$ which determines whether there is weak or strong fragmentation. The idea is that one takes the set of generated samples, and categorizes them based on Hilbert space fragment and global $U(1)$ charge. This procedure requires constructing an update protocol that is ergodic and satisifies detailed balance, and finding a method to identify (and label) the fragment that a given configuration belongs to.

Motivated by the examples in Figs.~\ref{fig:app_rhombus_vertex} and \ref{fig:app_Lieb_vertex}, we will consider a particular class of vertex models. These have coordination number $z$---in the typical case of $d$-dimensional hypercubic lattices, this will be $z=2d$. Hence vertices will have $z$ legs. Sites can have links on these legs, and there will be no constraint on the number of links allowed (unlike in ice-rule models). Vertices can also have additional decorations, though as explained below, this will only occur for linkless (ghost) vertices.

There are $2^z$ \emph{standard} vertices which simply correspond to the combinations of links emanating from an undecorated site. Any vertex that doesn't have links is denoted a \emph{ghost} vertex. There are $g$ types of ghost vertices, which includes the linkless standard vertex. They are distinguished by various decorations on the site. For example, $g=2$ for the 17-vertex model relevant for the Lieb lattice. Vertices with at least one link are \emph{solid}. The total number of vertex types is $v=2^z+g-1$. The allowed pairs of adjacent vertices are simply determined by checking if their shared legs are consistent about the presence of a link.

Consider two vertex configurations $c,c'$. We denote by $w_{cc'}$ the update probability corresponding to $c\rightarrow c'$ in a time step. This can be decomposed as
\begin{equation}
    w_{cc'}=u_{cc'}a_{cc'}
\end{equation}
where $u_{cc'}$ is the proposal probability and $a_{cc'}$ is the acceptance probability. Let's say we want to sample configurations according to some weight $e^{\mu f_c}$, where $f_c$ is some function of $c$ which will typically be the particle number (in which case $\mu$ is the chemical potential). A sufficient condition for convergence to this distribution is, on top of ergodicity, the property of detailed balance
\begin{equation}
    \frac{w_{cc'}}{w_{c'c}}=e^{\mu(f_{c'}-f_c)}=e^{\mu \Delta f},\quad \Delta f=f_{c'}-f_c.
\end{equation}
The weights are introduced to bias towards different fillings, otherwise the MC would be dominated by states in the vicinity of the filling with maximum entropy. Several MC runs with different $\mu$ are required to generate a sufficient number of samples for each particle number.

We now construct a proposal protocol that is symmetric. Once we have this, detailed balance can be easily achieved by choosing the acceptance probabilities as in the Metropolis or heatbath method. For a vertex model with only standard vertices, there is a straightforward proposal protocol that is symmetric. First we pick a site at random. Then we change (or more precisely, propose to change) this vertex to any other vertex with uniform probability $\frac{1}{v-1}$. This might violate the compatibility with the $z$ adjacent sites, but we can simply correct those vertices in a deterministic manner, with the condition that all other links are unchanged. In other words, at each step, only the vertices on a \emph{star} may change.

Complications arise with extra ghost vertices $(g>1)$. To see this, consider a simple 1d example with vertices $\{0,0^*,1_\text{W},1_\text{E},2\}$, which has two ghost vertices. Consider changing the central vertex of the star $1_\text{E}22$ to $1_\text{E}$. There are two possible corrections---we could produce either $01_\text{E}2$ or $0^*1_\text{E}2$ as outcomes. We cannot just decide to determinstically pick one, because both $01_\text{E}2$ and $0^*1_\text{E}2$ can flip to $1_\text{E}22$. 

Returning to the general case, here is one solution for a symmetric proposal protocol:
\begin{enumerate}
    \item Pick a site $x$ at random. Say its vertex is $a$. Consider the star of vertices $\{a,b_0,b_1\ldots,b_{z-1}\}$ which includes the vertices of sites adjacent to $x$.
    \item Pick a new different vertex $a'$ uniformly with probability $\frac{1}{v-1}$. This will be our candidate replacement for $a$, but as discussed below, we will not necessarily propose this update. By considering the compatibility of $a'$ with its neighbours, we obtain two integers $n_{\text{gs}},n_{\text{sg}}$. $n_{\text{gs}}$ counts the number of ghost vertices which are forced to become solid. $n_{\text{sg}}$ counts the number of solid vertices which need to become ghosts.
    \item The next step now bifurcates into two types: successful changes and failed changes.
    \begin{itemize}
        \item With probability $\frac{1}{g^{n_\text{gs}}}$, we end up successfully changing (i.e.~a successful proposal) $a\rightarrow a'$. Now there are still $g^{n_\text{sg}}$ new ghost vertices to assign. For each choice, we uniformly assign a multiplicative probability factor of $\frac{1}{g^{n_\text{sg}}}$.
        \item The alternative is to fail the change and leave $a$ alone. However, for the $n_{\text{gs}}$ vertices that would have been solidified if $a\rightarrow a'$, we assign a new distinct set of ghost vertices. There are $g^{n_\text{gs}}-1$ choices (since we exclude the original configuration of ghosts), and each choice has uniform probability $\frac{1}{g^{n_\text{gs}}}$.
    \end{itemize}
\end{enumerate}
To summarize, given a star configuration, there is a probability $\frac{1}{v-1}\frac{1}{g^{n_\text{gs}+n_\text{sg}}}$ for each outcome with a new central vertex $a'$ and choice of new ghosts, and probability $\frac{1}{v-1}\frac{1}{g^n_\text{gs}}$ for each outcome with a failed change $a\rightarrow a'$ but a different choice of ghosts on vertices that would have solidified for $a'$. It is straightforward to see that the proposal probability $u_{cc'}$ is symmetric. Say $c$ and $c'$ can be directly connected by an update. We can go in both directions for successful and failed changes. For a failed change, it is clear that the factor $\frac{1}{v-1}\frac{1}{g^{n_\text{gs}}}$ is the same in both directions for each `change channel' $a'$. For a successful change, the roles of $n_\text{gs}$ and $n_\text{sg}$ are reversed in the two directions, but the probability $\frac{1}{v-1}\frac{1}{g^{n_\text{gs}+n_\text{sg}}}$ remains the same. A good sanity check that the MC is generating samples sensibly is to compare the distribution function (as a function of particle number) with the tensor methods using the same chemical potential $\mu$. This protocol is clearly ergodic. To obtain any desired legal configuration, one can first change all the vertices on one sublattice to the target configuration. Then one changes vertices the other sublattice, while passing various probability checks to prevent altering the first sublattice.

\section{Blockades on other lattices}
\begin{figure}
    \centering
    \includegraphics[width=.5\textwidth]{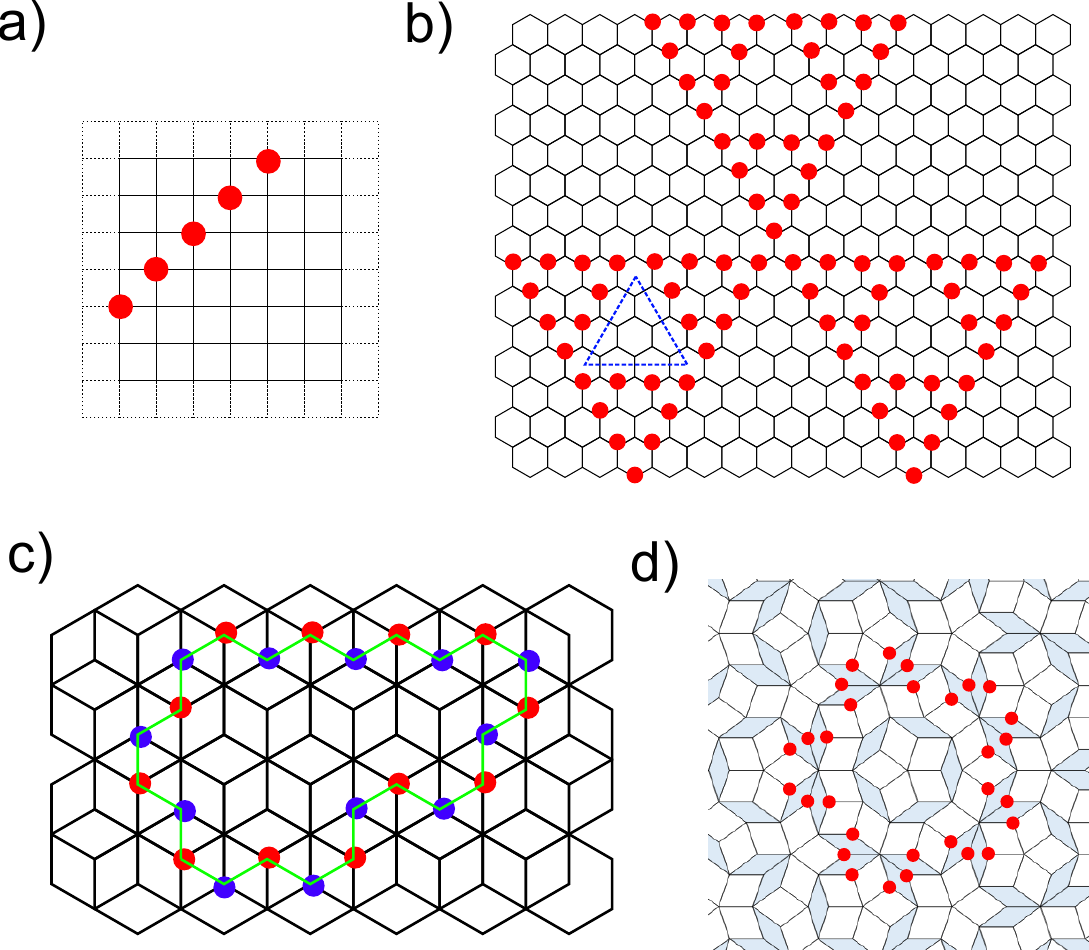}
    \caption{Example blockades for the a) square lattice, b) honeycomb lattice, c) dice lattice, d) P3 Penrose tilings.}
    \label{fig:app:other_lattices}
\end{figure}

In Fig.~\ref{fig:app:other_lattices}, we show example blockades on some other lattices. For the square lattice (Fig.~\ref{fig:app:other_lattices}a), a diagonal line of particles is a blockade. Note that this must extend across the system size, so there is no notion of a `local' blockade. In the thermodynamic limit, the minimum filling for fragmentation to occurs tends to zero. However, this system is likely weakly fragmented for all fillings for $N\rightarrow\infty$, similar to the square lattice transverse field Ising model at strong coupling. Another perspective is to start from the maximally filled sublattice-polarized state, and empty out rectangles (with sides parallel to the diagonals) to form bubbles. A similar situation occurs for the honeycomb lattice (Fig.~\ref{fig:app:other_lattices}b), where the minimal blockade is a Sierpinski triangle. Hence the minimum filling for fragmentation is zero, but the system is likely weakly fragmented. Another perspective is to start from the sublattice polarized state and empty out triangular regions of particles to form bubbles. The dice lattice (Fig.~\ref{fig:app:other_lattices}c), like the SLL, is capable of forming local blockade loops. These consist of occupied three-fold coordinated sites on a loop. The P3 Penrose tiling (Fig.~\ref{fig:app:other_lattices}d) also accommodates local loop blockades. The square Lieb lattice can be generalized to higher dimensions $d$, where Lieb sites are placed on the links of a hypercube. The blockades are still loops, which can now also form non-trivial knots. Conserved subregion charges (the analog of the membrane operators for $d=2$) occur when multiple blockades form a closed hypersurface of dimension $d-1$. 

\section{Details on exact enumeration and diagonalization}

In order to fully characterize the fragmentation on finite systems, we employ an exact enumeration technique on the basis elements of the Hilbert space in a given symmetry sector (i.e. at a given filling of fermions) on finite systems. Two basis elements (Fock states) belong to the same fragment if and only if they are connected by the Hamiltonian in Eq.~\ref{eq:H}. Since density-density interactions are diagonal in the occupation number basis, the only relevant terms are hoppings $\sim \hat{d}^\dagger_i\hat{d}_j$ subject to the hardcore constraint. This can be mapped to the problem of finding connected sub-graphs, where basis elements form the nodes and constrained hoppings the edges of the total graph. Starting from a seed node, all adjacent ones, that have not yet been visited, are added to the fragment and marked as visited. These act again as seed nodes for the next recursion until the fragment is exhausted and a new search is started from a node not yet belonging to any fragment. Because of its linear-in-depth memory requirements, the graph is traversed via a depth-first-search.

For the diagonalization, we explicitly construct the many-body Hamiltonian for a given fragment by first performing EE and subsequently building the matrix in the basis of product states belonging to that fragment. By definition, all matrix elements may only act within a given fragment which, in the case of strong fragmentation, reduces the computational complexity per diagonalization dramatically. In the strongly fragmented case of the rhombus chain with $L=12$ unit cells ($N_s=36$) at $N_p=13$ ($\nu\simeq 0.36$), the size of the largest fragment is $\mathcal{O}(10^4)$, making it possible to obtain the full spectrum whilst providing sufficient statistics. Fig.~\ref{fig:thermalization}b encompasses 6321 eigenvalues, while the full symmetry sector has a dimension of roughly 4.9 million.
We set the scale of the hopping amplitudes to $t=1$ and randomly choose their relative strengths in the uniform interval $[0.8,1.2]$. The same kind of randomization is applied to the interaction couplings. The onsite potential is chosen to be in $[-0.1,0.1]$.
The eigenvalues obtained from the diagonalization are used to construct the spectrum of level spacing ratios as outlined in the main text. The functional form of the Poisson probability distribution on the interval $[0,1]$ is $P(r)=\frac{2}{(1+r)^2}$  whilst for GOE we have $P(r)=\frac{27}{4} \frac{r+r^2}{(1+r+r^2)^{3/2}}$ \cite{Atas2013}.
It should be noted that not all fragments display a transition as clear as the one displayed in Fig.~\ref{fig:thermalization}b under the introduction of $V_4$. This is due to the interplay of `blockade thickness' and interaction range, as well as finite size effects which blur the underlying distribution.

\section{Cluster expansions at low filling}
We  prove that at low particle density, hardcore fermions on the Lieb lattice show strong Hilbert space fragmentation. While we consider the Lieb lattice for specificity, we will see that our results apply to all graphs with an extensive number of elementary blockade configurations. Different fragments of the Hilbert space correspond to different frozen configurations of  closed loops of the square lattice. Strong fragmentation corresponds to $D_{\rm max}/D _{\rm sym}\rightarrow 0$ in the thermodynamic limit. The starting assumption is that at the low particle densities, which is the focus of this section, the largest fragment corresponds to configurations with no frozen loops.
As mentioned in the main text, $D_{\rm sym}$  corresponds to the parition function $Z(\nu)$ of hardcore particles on the Lieb lattice, while $D_{\rm max}$ corresponds to the partition function $\overline{Z}(\nu)$ of hardcore particles with no blockade configurations (closed loops on the underlying square lattice).
We start with the related partition functions $Z(w)$ and $\overline{Z} (w)$ where particles have been endowed with a fugacity $w$. We use the Mayer cluster expansion~\cite{Mayer,FordUhlenbec,FriedliVelenik} to compute $\log(Z)$ and $\log\overline{Z}$ perturbatively in $w$. Inverting a  similar series expansion for the filling $\nu$ as a power series in $w$ allows us to compute $D_{\rm max}/D_{\rm sym}$ as a function of $\nu$ and show that it vanishes in the thermodynamic limit for small $\nu$.

\subsubsection{Statement and results}
To maintain the linearity of the presentation, here we state the cluster expansion for hardcore models  and follow it up with the consequent results for fragmentation. We present a derivation in the next subsection. We first introduce some terminology. A graph $G=(V,E)$ is given by a set of vertices $V$ and a set $E$ of edges or vertex pairs ($E \subseteq V \times V$). A subgraph $K \in G$ has vertex and edge sets which are subsets of $V$ and $G$ respectively. A subset $U\subseteq V$ of vertices ``induces" a subgraph $K \in G$ which has all vertices in $U$ and all edges $G$ which are incident on vertices in $U$. A ``spanning" subgraph $K\in G$ has the same set of vertices as the parent graph $G$. A sequence of edges $\{ (v_1,v_2),(v_2,v_3) \ldots (v_{n-1},v_n)\}$ is called a ``path" connecting the vertices $v_1$ and $v_n$. A graph $G$ is connected if every pair of vertices in it are connected by a path. A ``complete" subgraph has an edge between all pairs of edges in it.

 Consider the set $V$ of vertices of the Lieb lattice (which we will also consider as a graph $G=(V,E)$), and a subset $I \in V$ of \emph{compatible} (with the hardcore constraint) vertices occupied by hardcore particles. The set of vertices $I$ is compatible if every pair of vertices $(v_i,v_j)$ in $I$ are compatible, \textit{i.e.},  $(v_i,v_j) \notin E$. The partition function of particles with activity $w$ is given by :

\begin{align}
  Z(w) = \sum_{\substack{I \in V \\ \mathrm{compatible}}} w^{|I|}
\label{eq:partition_hardcore}
\end{align}

\begin{figure}[t]
\includegraphics[width=\textwidth]{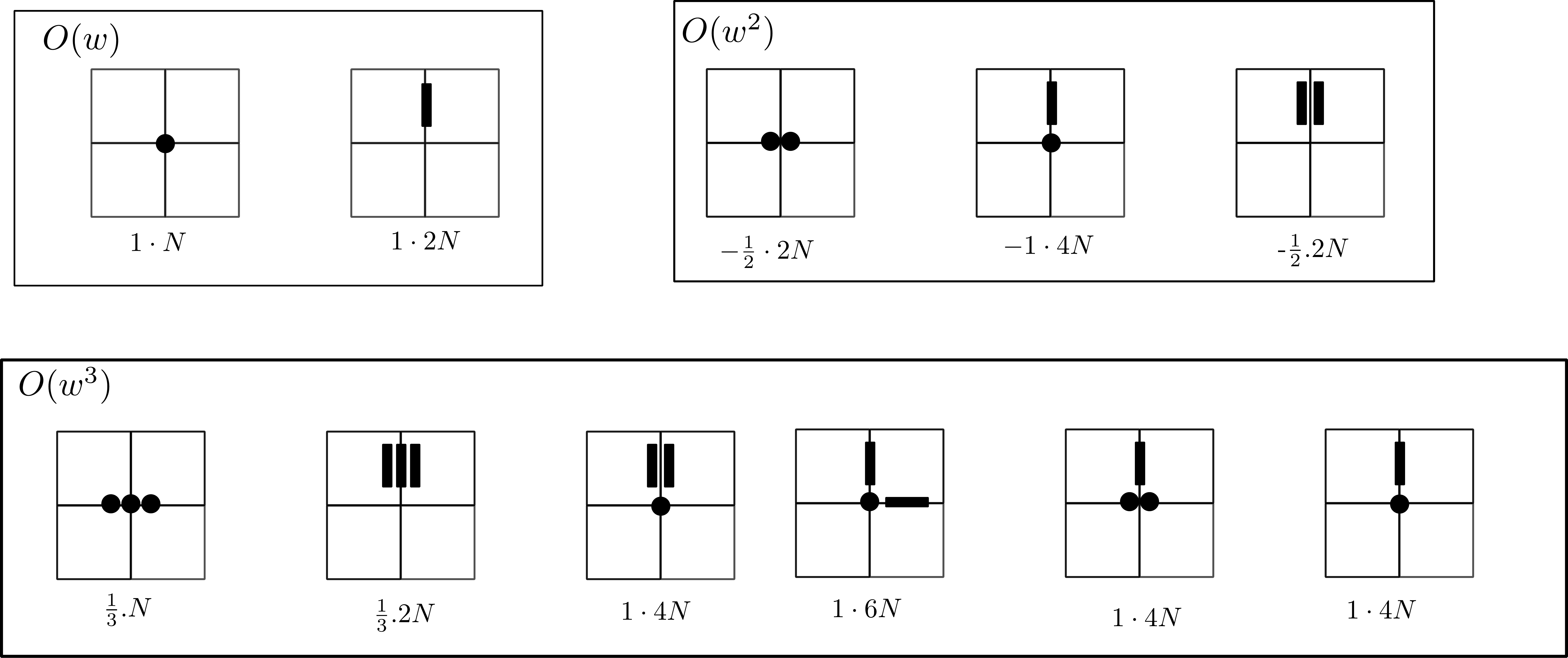}
\caption{\label{fig:diagrams}Each diagram shows a ``cluster" $X$ contributing to Eq.~\eqref{eq:cluster_expansion_final}, the numbers show the weight $a(X)$ and the number of such collections of the same weight obtained by lattice symmetries on a Lieb lattice of $N$ }
\end{figure}

The cluster expansion,  is a power series expansion of $\log Z$ in powers of $w$. The terms of this power series can be organised by connected objects on the graph $G$ called \emph{clusters}. A cluster $X$ is an unordered collection of vertices from the set $V$ with \emph{repetitions} allowed, such that the subgraph induced by $X$ on $G$ is connected. 
The cluster expansion is now be expressed as a sum over such clusters :
\begin{align}
  \log Z(w)  =\sum_{\substack{X \\ \mathrm{clusters}} } a(X) w^{|X|}\label{eq:cluster_expansion_final}
\end{align}
To write down an expression for the coefficient $a(X)$, we need to define two more quantities for the cluster $X$. First, we denote  the
number of occurrences of the vertex $v_i$ in $X$ by $n_X(i)$. Second, we define the graph $H(X)$:   we start with the connected subgraph induced on $G$ by vertices in $X$ and then replace each vertex $v_i$ by a complete graph on $n_X(i)$ vertices.
The coefficient $a(X)$ is now given by a  sum of connected, spanning subgraphs of $H(X)$.
\begin{align}
a(X) = \frac{1}{\prod_{i}n_X(i)!} \sum_{\substack{K\subseteq H(X) \\ \mathrm{conn., spann.}}}   (-1)^{|E_K|}.
\end{align}

For the Lieb lattice, we display the  clusters $X$ and their contributions to the cluster expansion for the first three orders in $w$ in Fig.~\ref{fig:diagrams}. Collecting all terms together, we have for the Lieb lattice with $N$ unit cells,
\begin{align}
\log(Z(w)) = 3Nw -\frac{11}{2}Nw^2 + 17Nw^3 - 62\frac{3}{4} Nw^4 +O(w^5)
\label{eq:logZ_expression}
\end{align}
The density of particles per unit cell, or filling fraction, is given by :
\begin{align}
\nu=w \dv{w}(\log Z(w)) = 3Nw -11Nw^2 + 51Nw^3 - 251 Nw^4 +O(w^5)
\label{eq:nu_expression}
\end{align}
While computing all terms by hand up to the fourth order is already laborious, algorithms to count polyominoes~\cite{Redelmeier} can be easily adapted to compute terms up to a few more orders. 

Now we shift focus to  $\overline{Z}$, the partition function of a related model where no frozen loops are allowed. This corresponds to the largest fragment at low particle densities. To state the expansion for this case, we need to introduce generalised clusters $\overline{X}$ which are subgraphs where all vertex pairs have an ``generalised connection" to each other. Two vertices have a generalised connection if they are connected to each other by a path, or if both vertices are a part of the same elementary blockade configuration (or square lattice loop, to be specific to the Lieb lattice).
 $\log \overline{Z}$ now has a similar cluster expansion over generalised clusters $\overline{X}$ :
\begin{align}
  \log \overline{Z}(w)  =\sum_{\substack{\overline{X} \\ \mathrm{gen. clust}} } a(\overline{X}) \prod_{v_i\in X}w(v_i). \label{eq:cluster_expansion_final2}
\end{align}

The coefficient is now given as a sum of ``generally connected" spanning subgraphs of $H(\overline{X})$ 
\begin{align}
a(\overline{X}) = \frac{1}{ \prod_{i}n_{\overline{X}}(i)!}  \sum_{\substack{K\subseteq H(X) \\ {\text{gen. conn. spann.}}}}  (-1)^{|E_K|} (1)^{|L_k|}.
\end{align}
We have introduced the $|L_k|$ to denote the number of square lattice loops in the subgraph $K$.
It is not hard to see that generalised clusters start differing from clusters at $O(w^4)$ where the smallest loops come into play, 
\begin{align}
  \log \overline{Z} = \log Z - N w^4 +O(w^5).
  \label{eq:logZbar_expression}
\end{align}
Inverting Eq.~\eqref{eq:nu_expression} to get $\nu = w/3 + O(w^2)$, we obtain
\begin{align}
  \frac{\overline{Z}}{Z}=\frac{D_{\mathrm{max}}}{D}\sim \exp(-\nu^4/84).
\end{align}
Therefore we have shown that if the particle numbers are large enough so that the equivalence of ensembles continue to hold,  hardcore fermions on the Lieb lattice remains strongly fragmented in the limit of low particle density. We emphasize that the $\exp(-\nu^4/84)$ behaviour emerges in the low density limit, but for a thermodynamically large number of particles. This is likely not the regime explored by the exact enumeration and Monte Carlo calculations presented in the main text.
We see that for any lattice where there are an extensive, $O(N)$ number of elementary blockade configurations of size $s$, this result would continue to hold with $D_{\rm max}/ D_{\rm sym
} \sim \exp(-c\nu^s N)$.

\subsubsection{Derivations of cluster representation of the free energies}

Consider a graph $G=(V,E)$, with the vertices hosting hardcore particles.
To formulate the cluster expansion, we write the partition function of hardcore particles as an unrestricted sum over an ordered list of vertices $\{v_1,v_2 \cdots v_n\}$ from $V$ with repetitions allowed:
\begin{align}
Z(w) = 1+\sum_{n}\frac{1}{n!}\sum_{v_1\in V}\sum_{v_2\in V}\ldots \sum_{v_n\in V} w^n \prod_{i,j\in[1,n] } \delta(v_i,v_j).
\label{eq:polymer_partition}
\end{align}
The factor $\delta(v_i,v_j)$ is $0$ if the vertices $v_i$ and $v_j$ are ``incompatible" (prohibited by hardcore conditoins or  $v_i=v_j$), otherwise it is $1$.   Now we use the `Mayer trick' to write $\delta(v_i,v_j)=1+f_{ij}$, with $f_{ij}=-1$ for incompatible vertices and $f_{ij}=0$ for compatible vertices; this allows us to rewrite Eq.~\eqref{eq:polymer_partition} as
\begin{align}
Z(w) = 1+\sum_{n}\frac{1}{n!}\sum_{v_1}\sum_{v_2}\ldots \sum_{v_n} w^n \prod_{i,j\in[1,n] } (1+f_{ij}).
\label{eq:mayer_trick}
\end{align}
Now consider a graph $G_n$ with a set of vertices $V_n=\prod^n_i V$  and an edge set $E_n$ corresponding to an edge between each pair of vertices in $V_n$; in other words $G_n$ is a \emph{complete} graph on the set of vertices $V_n$. What the Mayer trick allows us to do is to write the partition function as a sum over
subgraphs (connected and otherwise) of $G_n$
\begin{align}
Z(w)  =\sum_{n}\frac{1}{n!} w^n\sum_{G\in G_n}  \prod_{(ij)\in G} f_{ij}.
\label{eq:mayer_trick2}
\end{align}

The first Mayer theorem~\cite{FordUhlenbec} essentially states that only connected subgraphs contribute to $\log Z$
\begin{align}
\log Z(w)  =\sum_{n}\frac{1}{n!} w^n\sum_{\substack{G\subseteq G_n \\ \mathrm{connected}}}  \prod_{(ij)\in G} f_{ij}.
\label{eq:cluster_expansion}
\end{align}
There is a further simplification for our case of hardcore interactions. Consider the subgraph $G \subseteq G_n$ in Eq.~\eqref{eq:cluster_expansion} above to have a vertex set $U=\{v_1,v_2 \cdots v_n\}$. Such a vertex set is called decomposable if it can be decomposed in two sets $U_a$ an $U_b$, such that each vertex in $U_a$ is compatible with all other vertices in $U_b$. Since $f_{ij}=0$ for such compatible vertex pairs, subgraphs $G\subseteq G_n$  with decomposable vertex sets do not contribute to the sum in Eq.~\eqref{eq:cluster_expansion}. This brings us to clusters. To facilitate computations and collect terms in Eq.~\eqref{eq:cluster_expansion}, we introduce a cluster $X$, i.e.~an unordered list of non-decomposable vertices of the original graph with  repetitions. If a vertex $v_i$ occurs $n_X(i)$ times in the collection $X$, each such collection corresponds to $\frac{n!}{\prod_i n_X(i)!}$ connected subgraphs $G\in G_n$ in Eq.~\eqref{eq:cluster_expansion}. Further, we define $G(X)$  to be the graph obtained by starting with 
the subgraph of the Lieb lattice $G$ induced by the vertices in $X$, and then replacing each vertex $v_i$ by
a complete graph on $n_X(i)$ vertices.
Bringing it all together, and using $f_{ij}=-1$ for incompatible vertex pairs, we have
\begin{align}
\log(Z(w))  =\sum_{X}w^{|X|}\underbrace{\frac{1}{\prod_{i}n_X(i)!} \sum_{\substack{K\subseteq H(X) \\ \mathrm{conn. spann.}}}  \prod_{(ij)\in G} (-1)^{|E_K|}}_{a(X)}.
\label{eq:cluster_expansion_final}
\end{align}

Now we present the extension for $\log ( \overline Z)$. The parition function is now given by

\begin{align}
Z(w) = 1+\sum_{n}\frac{1}{n!}\sum_{v_1\in V}\sum_{v_2\in V}\ldots \sum_{v_n\in V} w^n \prod_{i,j\in[1,n] } \delta(v_i,v_j) \prod _{\substack{ L  \\ {\rm loop}}} \delta_L.
\label{eq:polymer_partition_bar}
\end{align}
The last product now runs over all loops of the underlying square lattice, and $\delta_L$ evaluates to $0$ if all vertices on the loop $L$ are occupied by vertices in the set $\{v_1 \ldots v_n\}$. Applying the Mayer trick now gives us 
\begin{align}
\overline{Z}(w)  =\sum_{n}\frac{1}{n!} w^n\sum_{G\in G_n}  \prod_{(ij)\in G} f_{ij} \prod_{L\in G} f_{L},
\label{eq:mayer_trick2_bar}
\end{align}
where the last product is over all loops among the vertices of the subgraph $G$. Crucially, the first Mayer theorem 
continues to hold, and $\log \overline{Z}(w)$ can be expressed as the sum over ``generally connected" graphs, where two vertices  have a generalised connection if they are connected by a path, or are a part of the same square lattice loop. For other lattices, generalised connections can be by extending connections to two vertices being a part of the same elementary blockade configuration.

\section{Mazur bounds on charge autocorrelation functions}
Consider the infinite-temperature long-time average of the dynamical autocorrelation function of some observable $\hat{O}(t)=e^{i\hat{H}t}\hat{O}e^{-i\hat{H}t}$
\begin{equation}
    C_O=\lim_{\tau\rightarrow\infty}\frac{1}{\tau}\int_0^\tau dt \langle \hat{O}(t) \hat{O}(0)\rangle\geq M_O,
\end{equation}
where the expectation value is taken over the infinite-temperature canonical or grand canonical ensemble. The Mazur bound provides a lower bound $M_O\leq C_O$ by using the symmetries of the Hamiltonian. In Ref.~\cite{Moudgalya2022commutant}, this was applied to the commutant algebra that characterizes families of fragmented systems. Define the operator overlap $(A|B)=\langle A^\dagger B\rangle=\frac{1}{D}\text{Tr}(A^\dagger B)$, where $D$ is total Hilbert space dimension under consideration. For a set of orthogonal conserved quantities $Q_\alpha$, the Mazur bound takes a particularly simple form
\begin{equation}
    M_O(\{\hat{Q}_\alpha\})=\sum_{\alpha}\frac{(O|Q_\alpha)(Q_\alpha|O)}{
    (Q_\alpha|Q_\alpha)}.
\end{equation}
A possible choice for the $Q_\alpha$ is the set of projectors onto the Hilbert space fragments.

We now specialize to the hardcore extended Hubbard models discussed in the main text. We distinguish between the grand canonical (no subscript) and canonical (subscript $N_p$) ensembles, since particle number $\hat{N}_p$ is the only conventional symmetry in the problem. They have total Hilbert space dimensions $D_\text{tot}$ and $D_{\text{tot},N_p}$ respectively. Clearly $D_\text{tot}=\sum_{N_p}D_{\text{tot},N_p}$. The fixed-$N_p$ sectors are associated with projectors $\hat{Q}_{N_p}$ onto states with fixed particle number $N_p$. The Hilbert space shatters into fragments of dimension $D_{N_p,\alpha}$, where $\alpha$ runs over the fragments for a fixed particle number. They are associated with projectors $\hat{Q}_{N_p,\alpha}$. Consider an observable $\hat{O}$ that is diagonal in the Fock basis. We can define various infinite-temperature averages: \begin{itemize}
    \item $\langle O\rangle$ is the average of $\hat{O}$ over the entire Hilbert space (over all $N_p$ sectors).
    \item $\langle O\rangle_{N_p}$ is the average of $\hat{O}$ over a fixed $N_p$ sector.
    \item $\langle O\rangle_{N_p,\alpha}$ is the average of $\hat{O}$ over a single Hilbert space fragment $(N_p,\alpha)$.
\end{itemize}
For the choice of observable, we choose the particle density on a given site $j$, measured with respect to the average over the relevant ensemble. The constant offset is to make $\hat{O}$ traceless, so that the trivial symmetry $\mathbb{1}$ has a vanishing contribution to the Mazur bound. For the canonical ensemble, the operator is $\hat{O}=\hat{n}_j-\langle n_j\rangle_{N_p}$. For the grand canonical ensemble, the operator is $\hat{O}=\hat{n}_j-\langle n_j\rangle$. Note that the bound is expected to depend on the sublattice. In particular, sublattices which participate directly in blockades (such as the Lieb sites on the SLL) are expected to have a greater bound than those who do not (such as the square lattice sites on the SLL). These quantities are straightforwardly extracted from the exact enumeration of fragments, and the computation can be easily generalized to other diagonal observables since as non-local products of densities.

\subsubsection{Canonical ensemble}
Consider first the canonical ensemble with fixed $N_p$. The observable of interest is $\hat{O}=\hat{n}_j-\langle n_j\rangle_{N_p}$, where $\langle n_j\rangle_{N_p}$ is
\begin{equation}
    \langle n_j\rangle_{N_p}=\frac{1}{D_{\text{tot},N_p}}\text{Tr}_{N_p}\hat{n}_j=\frac{\sum_\alpha D_{N_p,\alpha}\langle n_j\rangle_{N_p,\alpha}}{D_{\text{tot},N_p}},
\end{equation}
and $\text{Tr}_{N_p}$ indicates a trace over the Hilbert space of fixed $N_p$. Total particle conservation alone does not contribute to the Mazur bound since we have subtracted the average density already, i.e.~$M_O(\hat{N}_p)=0$. If we consider the set of projectors $\{\hat{Q}_{N_p,\alpha}\}$ onto the fragments, we obtain the Mazur bound
\begin{equation}
M_O(\{\hat{Q}_{N_p,\alpha}\})=\frac{1}{D_{\text{tot},N_p}}\sum_\alpha D_{N_p,\alpha}\left(\langle n_j\rangle_{N_p,\alpha}-\langle n_j\rangle_{N_p}\right)^2.
\end{equation}

\subsubsection{Grand canonical ensemble}
Consider now the grand canonical ensemble which includes all particle number sectors. The observable of interest is $\hat{O}=\hat{n}_j-\langle n_j\rangle$, where $\langle n_j\rangle$ is
\begin{equation}
    \langle n_j\rangle=\frac{1}{D_\text{tot}}\text{Tr}\,\hat{n}_j=\frac{\sum_{N_p}\sum_\alpha D_{N_p,\alpha}\langle n_j\rangle_{N_p,\alpha}}{D_\text{tot}}.
\end{equation}
Because of the offset, the trivial symmetry $\mathbb{1}$ does not contribute to the Mazur bound, i.e.~$M_O(\mathbb{1})=0$. Considering just total particle conservation via the set of projectors $\{\hat{Q}_{N_p}\}$, we obtain the Mazur bound
\begin{equation}
    M_O(\{\hat{Q}_{N_p}\})=\frac{1}{D_\text{tot}}\sum_{N_p}D_{N_p}\left(\langle n_j\rangle_{N_p}-\langle n_j\rangle\right)^2.
\end{equation}
We can obtain a tighter bound by considering the set of all projectors onto Hilbert space fragments, leading to
\begin{equation}
M_O(\{\hat{Q}_{N_p,\alpha}\})=\frac{1}{D_{\text{tot}}}\sum_{N_p}\sum_\alpha D_{N_p,\alpha}\left(\langle n_j\rangle_{N_p,\alpha}-\langle n_j\rangle\right)^2.
\end{equation}

\subsubsection{Numerical results for the square Lieb lattice}

\begin{figure}[t]
    \centering
    \includegraphics[width=0.6\columnwidth]{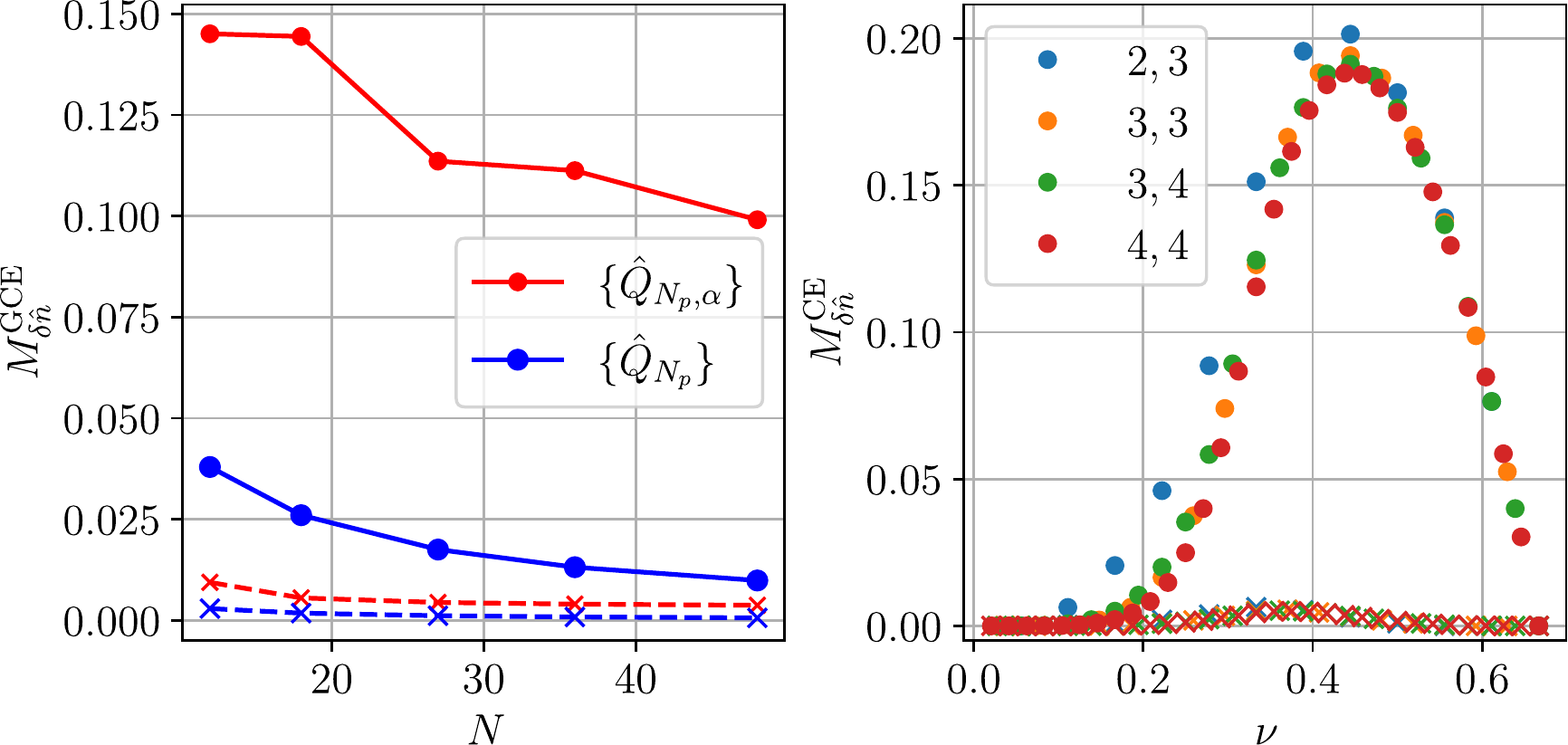}
    \caption{Mazur bounds for the dynamical autocorrelator of $\delta \hat{n}_j=\hat{n}_j-\langle n\rangle_j$ in the square Lieb lattice. Left: Red (blue) indicates bounds in the grand canonical ensemble, taking into account fragmentation (only $N_p$ conservation). Circles (crosses) indicate Lieb (square lattice) sites. Right: Bounds obtained using fragmentation in the canonical ensemble.}
    \label{fig:app:liebmazur}
\end{figure}

In the main text, we show results for the rhombus chain using exact enumeration. In Fig.~\ref{fig:app:liebmazur}, we show analogous results for the square Lieb lattice. Note that the grand canonical bound using the full fragmented structure still shows severe finite-size effects, owing to the fact that we use periodic boundary conditions and at least one of the side lengths is $\leq4$. 

\end{appendix}

\end{document}